\documentclass[aps,prl,twocolumn,groupedaddress,showpacs]{revtex4}
\usepackage{amsmath,amssymb,graphicx}

\begin{document}
\title{{Magnetization dependent current rectification in (Ga,Mn)As magnetic tunnel junctions}}

\author{Yoshiaki Hashimoto, Hiroaki Amano, Yasuhiro Iye, Shingo Katsumoto}
\affiliation{Institute for Solid State Physics, University of Tokyo, 5-1-5 Kashiwanoha, Kashiwa, Chiba 277-8581, Japan}
\begin{abstract}
We have found that the current rectification effect in triple layer (double barrier) (Ga,Mn)As magnetic tunnel junctions strongly depends on the magnetization alignment.
The direction as well as the amplitude of the rectification changes with the alignment,
which can be switched by bi-directional spin-injection with very small threshold currents.
A possible origin of the rectification is energy dependence of the density of states around
the Fermi level. Tunneling density of states in (Ga,Mn)As shows characteristic dip around zero-bias indicating 
formation of correlation gap, the asymmetry of which would be a potential source of the energy dependent
density of states.
\end{abstract}

\maketitle

Control of device characteristics through spin-degree of freedom is one of the expected
novel functionalities in semiconductor spintronics.
In metallic magnetic tunnel junctions (MTJs), rectification of microwaves
driven by ferromagnetic resonance (FMR) has been reported~\cite{Tulapurkar2005}.
This arises interests in the effect of ferromagnetism on the rectification
properties in diluted magnetic semiconductor (DMS) devices.
In strongly asymmetric structures such as Schottky or p-n junctions, the rectification is naturally
realized through steep electrostatic band-bending, which cannot be expected 
in unipolar MTJs ordinarily obtained in DMSs.
For example in the case of (Ga,Mn)As\cite{Ohno1996}, the material should be p-type for the appearance of the ferromagnetism\cite{Komori2003}, which is mediated by charge carrying holes.
Instead spin-polarization may thus be utilized to commutate alternative currents.
In this letter, we report such rectification effect in (Ga,Mn)As tri-layer MTJs, in which 
very small threshold current for spin-injection magnetization reversal~\cite{Watanabe2008}.
The rectification is sensitive to the alignment of the magnetizations and hence can be controlled through bi-directional current injection.

\begin{figure}
\hfil\includegraphics[width=0.9\linewidth]{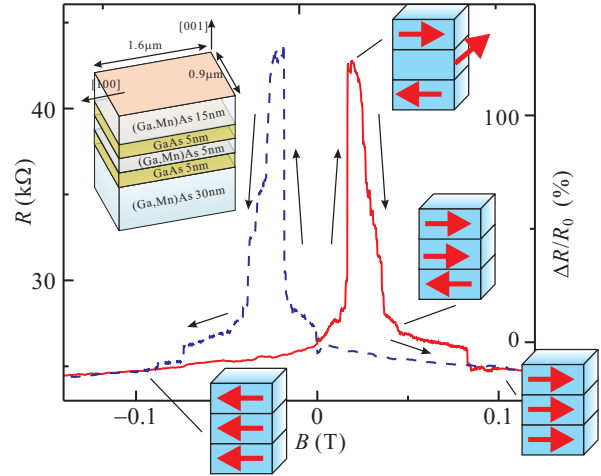}
\hfil
\caption{Typical TMR of a tri-layer (Ga,Mn)As MTJ. The arrows along the curve show directions of resistance
variations with a field cycling. Stacked blocks with arrows schematically show rough hypothetical configurations of magnetization at some points of the field.
The inset is a cross sectional view of LTMBE-grown layered structure.}
\label{fig_tmr}
\end{figure}

The inset of Fig.\ref{fig_tmr} schematically shows the cross section of the layered structure grown by
standard low temperature molecular beam epitaxy (LTMBE) onto a (001) p$^+$(Zn-doped) GaAs substrate\cite{Shen1997}
with the growth temperature of 320C.
The middle ``free" layer is designed to be thinner than the other two to have smaller coercive force,
which turned out to be, in reality, comparable to that of the top layer as we will see later.
The ferromagnetic transition temperature estimated from
the appearance of hysteretic tunneling magnetoresistance (TMR) is about 40K.
With use of electron beam lithography, the film was wet-etched into mesas with 
1.4$\times$0.9 $(\mu{\rm m})^2$ rectangles along [100], which is an easy axis of the
in-plane magnetization though the anisotropy should be weakened through the stress relaxation\cite{Suda2010}.
Each device was attached to a coplanar waveguide (CPW) with [100] parallel to the external magnetic field.
Here the doped substrate is electrically bonded to the center strip and separately
ordinal wires for DC are connected. The AC and DC lines were cut by bias tees.
Due to the high impedance of the devices, the AC sources drove the device voltages and the rectified 
currents were measured through a current-voltage amplifier or calculated 
from the DC voltages along the device and the I-V characteristics.
The external magnetic field of 0.7T was applied along [100] during cooling from room temperature to 4.2K.

Figure \ref{fig_tmr} shows a typical TMR for a major magnetic field loop. The lineshape resembles
to those reported so far~\cite{Watanabe2008}, suggesting similarity in magnetization alignments,
of which a possible set are illustrated for typical values of magnetic field.
Actually there should be several variations in the alignment around those simplified ones as can be seen in
small step structures.
This is probably due to the reduction in the in-plane crystallographic anisotropy
through relaxation of biaxial strain with micro-fabrication\cite{Suda2010}.

\begin{figure}
\includegraphics[width=\linewidth]{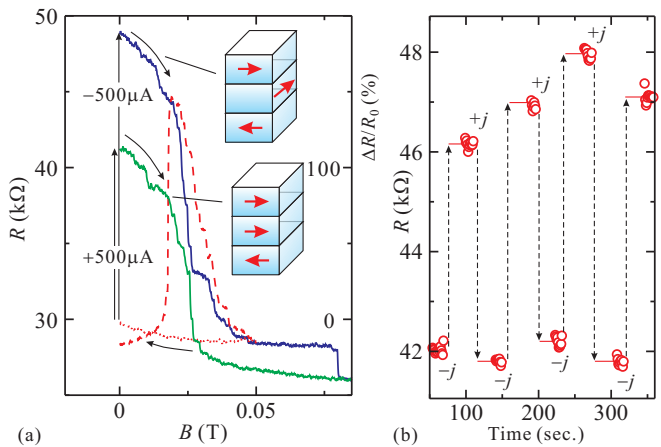}
\caption{(a) Magnetic minor loop for the MTJ (dotted curve), the major loop of which is shown in Fig.\ref{fig_tmr} 
and resistance jumps after current injection from the substrate to the upper electrode ($-$500$\mu$A) and
that with inverse direction (+500$\mu$A).
(b) Resistance switching between the two resistance states with the alternative current pulses ($\pm j$).
Missing of resistance data after the injections is due to dead-time of the resistance bridge.
}
\label{fig_rjump}
\end{figure}
Note that smallest coercive force is assumed in the the top layer in the illustrated alignments
in order for natural interpretation of the response to current injection.
This clearly appears in the response to the spin injection by current pulses.
A sequence consists of a minor field loop 0T $\rightarrow$ 0.06T $\rightarrow$ 0T and a current injection at zero field is adopted in Fig.\ref{fig_rjump}(a).
As shown in Fig.\ref{fig_rjump}(a), the junction resistance jumps up to the branch of the highest resistance
when the electric current of 500$\mu$A flows from the substrate to the upper electrode (we define this direction as `negative' current). The pulse width can be as short as 3ms, which is limited by wiring in the
cryostat.
A resistance leap with smaller increment is observed also for positive current pulse.
To assign these states of resistance to the alignments of magnetization, we need to assume

The switching between these two is reversible, {\it i.e.}, as shown in Fig.\ref{fig_rjump}(b),
the state flip-flops between them with alternating current injections, which can be explained
with assuming conditions as 90 degree rotation of the free layer with injected spin torques.

The current-voltage (I-V) characteristics were strongly non-linear, that is,
the differential resistance is reduced with the source-drain voltage as shown in Fig.3(a).
When the junction is driven by an AC voltage, time-averaged resistance is hence
reduced, and this can be utilized for estimation of the AC amplitude.
We here eliminated the possibility of heating under the AC voltages
by measuring the TMR and hence the magnetic coercive force, 
which is found to be sensitive to the temperature.
A kind of conductance, $G_{\rm r}=$(rectified DC current)/(AC drive voltage amplitude), can therefore serve
as a measure of the rectification.

\begin{figure}
\includegraphics[width=\linewidth]{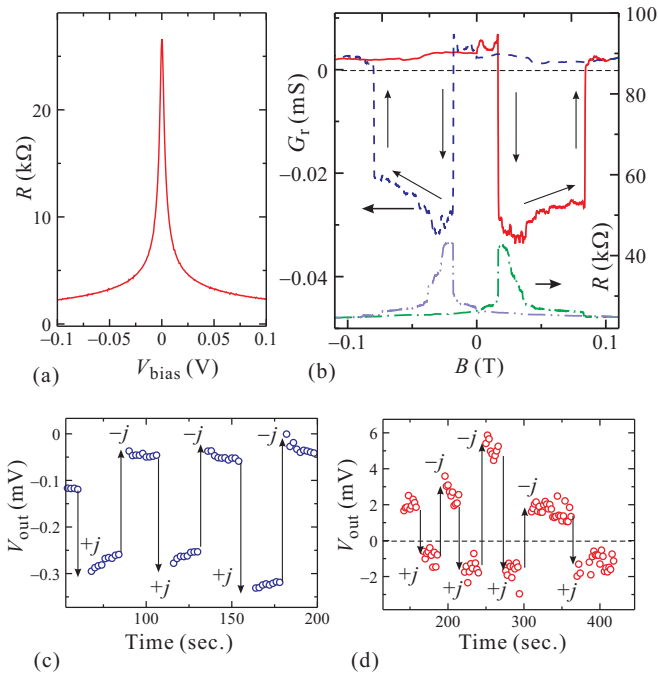}
\caption{(a) Bias voltage variation of differential resistance in the device, the TMR of which is shown in Fig.\ref{fig_tmr} ($B=0$).
(b) Rectification `conductance' of the junction (see the text for the definition) as a function of magnetic field
for 1.4GHz AC voltage with the amplitude of 1.7mV.
For reference, the TMR is also plotted.
(c) Switching of the rectified voltage output (the voltage terminals are open so the circuit resistance is
that of the junction itself) for current injections of 500$\mu$A, 1sec. with alternative directions.
(d) Another example of rectification switching for 1kHz AC drive.
The direction of the rectification is switched by the current injections.
}
\label{fig_vout}
\end{figure}

In spite of the almost symmetric I-V characteristics, a clear rectification voltage appears with the application
of AC voltages.
Figure \ref{fig_vout}(a) shows $G_{\rm r}$ as a function of external magnetic field at the AC frequency $f$ of
1.4GHz.
$G_{\rm r}$ has a small positive value for parallel alignments and 
large negative outputs appear for alignments with anti-parallel or perpendicular junctions.
Clear correspondence between $G_{\rm r}$ and the TMR is observed though it is not monotonic.
Because $G_{\rm r}$ is sensitive to the magnetization alignment, it also can be controlled with spin-injection.
Figure \ref{fig_vout}(b) displays flip-flop switching of the rectification conductance. The conditions can also be tuned
so that the direction of the rectification can be switched with spin-injection.
Figure \ref{fig_vout}(c) demonstrates such reversal of rectification direction by current (spin) injection
at zero field in another device.

Henceforth we discuss possible origin of the rectification.
The fact that the sign of output changes with the configuration of magnetization
eliminates the possibility that the source is zero-point shift
in the I-V characteristics due to some imbalance in the 
circuit including thermoelectric voltages in cryogenic wires, which only 
modifies the amplitude.
We have measured frequency dependence of $G_{\rm r}$ from 1GHz to 8GHz  (the limit of our microwave source) and found it very weak.
The spin torque diode effect is ruled out because it is based on FMR and thus should have strong frequency dependence.

We look for a possible explanation in the simplest model of magnetic tunnel junctions (MTJs) by Julliere\cite{Julliere}, in which
the conductance of an MTJ is a function of the majority and minority subbands density of states
in the electrodes.
A slight modification here is energy ($\epsilon$) and electrode ($j=1,2$) dependent density of states $f_j(\epsilon)$ and $g_j(\epsilon)$ for the majority and the minority spin subbands respectively.
We take the Fermi energy $\epsilon_{\rm F}$ as zero and the current differences
$\varDelta J_{\rm a,p}$ for voltages $\pm V$ are proportional to sum of convolutions as
\begin{align}
\Delta J_{\rm a}(V)&\equiv J_{\rm a}(V)+J_{\rm a}(-V)\notag\\
&\propto
\int_0^{eV}d\epsilon
\left\{
f_1(\epsilon-eV)g_2(\epsilon)+g_1(\epsilon-eV)f_2(\epsilon)
\right.\notag\\
&\;\;\;\left.
-f_2(\epsilon-eV)g_1(\epsilon)-g_2(\epsilon-eV)f_1(\epsilon)
\right\}
\label{eq_non_const_densityofstates_a}
\\
\varDelta J_{\rm p}&\propto
\int_0^{eV}d\epsilon\left\{f_1(\epsilon-eV)f_2(\epsilon)+g_1(\epsilon-eV)g_2(\epsilon)
\right.\notag\\
&\;\;\;\left.
-f_2(\epsilon-eV)f_1(\epsilon)-g_2(\epsilon-eV)g_1(\epsilon)
\right\}
\label{eq_non_const_densityofstates_p}
\end{align}
for anti-parallel and parallel configurations respectively.

It is apparent from \eqref{eq_non_const_densityofstates_a}, \eqref{eq_non_const_densityofstates_p} that
rectification stems from
the above mechanism only when 
the density of states are asymmetric to $\epsilon_{\rm F}$ and also between the electrodes.
In MTJ's with metal ferromagnets constant density of states approximation usually holds around $\epsilon_{\rm F}$
and the rectification does not appear.

To check whether the density of states in (Ga,Mn)As has such energy dependence,
we prepared a tunnel junction, which connects (Ga,Mn)As and p$^+$-GaAs layers with a single 
AlGaAs barrier layer (the inset of Fig.4(a)).
The film was cut into a 10$\times$10 ($\mu$m)$^2$ mesa and two metal electrodes were placed both on
the top and the bottom layers for four wire measurement.

Figure 4(b) shows the differential conductance of the junction as a function of bias voltage
in four wire measurement.
At the origin we observe a dip structure, which can be expressed as $\sim\epsilon^2$.
This probably due to so called Efros-Shklovskii (ES) gap, which originates from electron-electron
configuration interaction in disordered insulators.
The ES gap manifests that the hole states in (Ga,Mn)As is close to those in disordered insulators
rather than those in degenerate semiconductors.
However the gap cannot be the direct origin of the rectification because of the parabolic energy dependence
around $\epsilon_{\rm F}$, which is symmetric to $\epsilon_{\rm F}$.
Besides the gap structure we find significant asymmetry to zero-bias, which can cause the rectification
in combination with the difference between layers due to that in the thicknesses, etc.
Such asymmetry is more natural for the impurity band model than the degenerate semiconductor.

\begin{figure}
\includegraphics[width=\linewidth]{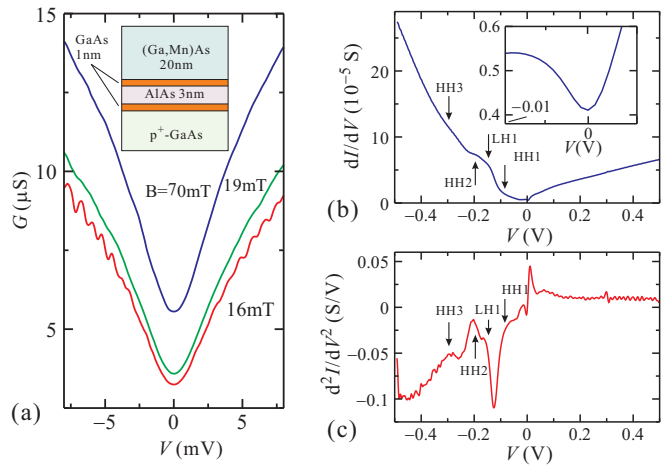}
\caption{(a) Conductance($G$)-voltage characteristics of a tri-layer MTJ for three different magnetic fields.
The inset shows a schematic cross sectional view of the layered structure.
(b) Differential conductance of the tunnel junction as a function of the bias voltage.
The inset is a blow up around the origin. Quantized level peak assignments are indicated by vertical arrows.
Because the quantization is along $z$-axis, the effective masses are taken as those for the bulk of GaAs.
(c) Numerical derivative of the data in (b) for the clarification of peak positions.
}
\label{fig_tunnel_conductance}
\end{figure}

Attention must be paid also to characteristic peaks and shoulders in negative bias.
These can be interpreted as the result of quantum confinement in the (Ga,Mn)As layer, which is put between
the internal (Al,Ga)As layer and the surface Schottky barrier\cite{Ohya2007}.
The voltage derivative of differential conductance is shown with a possible quantized level assignment
is shown in Fig.4(b).
For the assignment, we assume the effective thickness of the (Ga,Mn)As as 8nm, namely the Schottky depletion
width as 12nm. The triangular potential with 12nm width and 0.5eV height gives 300 $\Omega$/($\mu$m)$^2$
for the contact resistance, which is in reasonable agreement with that estimated from the difference between
the two-wire and the four-wire resistances (300-800 $\Omega$/($\mu$m)$^2$).
The result indicates that while the impurity band states are nearly localized, the valence band of matrix GaAs is kept comparatively ordered and coherent.
Also the result suggests the possibility of designing the rectifying characteristics because  these anomalies can be controlled through the thicknesses of the constituent layers.

In summary, we have found current rectification effect in tri-layer (Ga,Mn)As MTJs for AC voltages up to
8GHz. The rectification is strongly dependent on the alignment of the magnetizations and can be 
reproducibly switched by using current injection with very low threshold at zero field.
We have shown that the asymmetry of the density of states at the Fermi energy is a possible origin
of the rectification.

This work was supported by Grant-in-Aid for Scientific Research
and Special Coordination Funds for Promoting Science and
Technology.


\begin{thebibliography}{99}
\bibitem{Tulapurkar2005} A. A. Tulapurkar, Y. Suzuki, A. Fukushima, H. Kubota, H. Maehara, K. Tsunekawa, D. D. Djayaprawira, N. Watanabe, and S. Yuasa: Nature {\bf 438} (2005) 339.
\bibitem{Ohno1996} H. Ohno, A. Shen, F. Matsukura, A. Oiwa, A. Endo, S. Katsumoto, and Y. Iye: Appl. Phys. Lett.
{\bf 69} (1996) 363.
\bibitem{Komori2003} T. Komori, T. Ishikawa, T. Kuroda, J. Yoshino, F. Minami, and S. Koshihara:
\bibitem{Watanabe2008} M. Watanabe, J. Okabayashi, H. Toyao, T. Yamaguchi, and J. Yoshino: Appl. Phys. Lett. {\bf 92} (2008) 262.
\bibitem{Shen1997} A. Shen, and H. Ohno, F. Matsukura, Y. Sugawara, N. Akiba, T. Kuroiwa, A. Oiwa, A. Endo, S. Katsumoto, and Y. Iye: J. Crys. Growth. {\bf 175} (1997) 1069.
\bibitem{Suda2010} K. Suda, S. Kobayashi, J. Aoyama, and H. Munekata: IEEE trans. Magnetics {\bf 6} (2010) 2421.
\bibitem{Julliere} M. Julliere: Phys. Lett. A {\bf 54} (1975) 225.
\bibitem{Ohya2007} S. Ohya, K. Tkata, I. Muneta, P. N. Hai, and M. Tanaka: arXiv:1009.2235.
Phys. Rev. B {\bf 67} (2003) 115203.
\end{thebibliography}
\end{document}